# sp carbon chain interaction with silver nanoparticles probed by Surface Enhanced Raman Scattering

A. Lucotti[1], C. S. Casari[2], M. Tommasini[1], A. Li Bassi[2], D. Fazzi[1], V. Russo[2], M. Del Zoppo[1], C. Castiglioni[1], F. Cataldo[3], C. E. Bottani[2], G. Zerbi[1]

[1] Dipartimento di Chimica, Materiali e Ingegneria Chimica 'G. Natta' and
NEMAS - Center for NanoEngineered MAterials and Surfaces,
Politecnico di Milano, Piazza Leonardo da Vinci 32, I-20133 Milano, Italy

[2] Dipartimento di Energia and
NEMAS - Center for NanoEngineered MAterials and Surfaces,
Politecnico di Milano, via Ponzio 34/3, I-20133 Milano, Italy

[3] Actinium Chemical Research srl, via Casilina 1626/A, 00133 Roma, Italy and
INAF – Osservatorio Astrofisico di Catania, Via S. Sofia 78, 95123 Catania, Italy

**Abstract**

Surface Enhanced Raman Spectroscopy (SERS) is exploited here to investigate the interaction of isolated sp carbon chains (polyynes) in a methanol solution with silver nanoparticles. Hydrogen-terminated polyynes show a strong interaction with silver colloids used as the SERS active medium revealing a chemical SERS effect. SERS spectra after mixing polyynes with silver colloids show a noticeable time evolution. Experimental results, supported by density functional theory (DFT) calculations of the Raman modes, allow us to investigate the behaviour and stability of polyynes of different lengths and the overall sp conversion towards $sp^2$ phase.

**1. Introduction**

Linear carbon chains with sp hybridization represent one of the simplest one dimensional systems and have therefore attracted a great interest for many years [1, 2]. sp chains can display two types of carbon-carbon bonding: polyynes, chains with single-triple alternating bonds (…-C≡C-C≡C-…) and polycumulenes, chains with all double bonds (…=C=C=C=…). These linear forms of carbon are thought to be relevant in the initial stages of formation of fullerenes and nanotubes [3-5] and share the physics of Peierls distorsion and Kohn anomaly [6] with other polyconjugated systems such as polyacetylene [7], graphite and nanotubes [8, 9]. sp carbon has been produced by means of different physical and chemical techniques [10-17] and has been also observed in the core of multiwalled carbon nanotubes [18, 19]. Interesting transport properties have been predicted and measured [20, 21] and a strong non-linear optical response of finite linear carbon chains has been recently measured [22].

Raman spectroscopy is widely used for the investigation of sp carbon systems [23, 24, 25, 26] as well as for a number of carbon-based systems [7, 27]. In addition, Surface Enhanced Raman Scattering (SERS) can be exploited to achieve a high sensitivity in detecting small amounts of sp carbon. SERS may also be exploited to follow time dependent effects, thanks to the fast spectral recording allowed by the high signal enhancement. Therefore the study of vibrational properties of sp carbon chains through SERS spectroscopy is relevant for providing experimental data of these systems, as demonstrated in our recent works dealing with polyynes produced by the submerged arc discharge method [28, 29] and by other investigations of polyynes produced by laser ablation in liquids [30].

Here we exploit SERS to investigate time dependent effects when polyynes are interacting with a silver nanoparticle colloidal solution used as the SERS active medium. Hydrogen-terminated polyynes have a key role in the aggregation of silver colloids and show a high SERS signal [28, 30]. The overall SERS signal is affected by several effects such as Ag-molecule interaction, aggregation and sedimentation of aggregates of silver nanoparticles as well as molecular modifications due to the evolution of the sp chain population. Excluding spurious effects by proper normalization of the SERS spectra we observe the time evolution of polyynes interacting with silver nanoparticles revealing an overall tendency towards sp→$sp^2$ conversion in which shorter chains display a higher stability with respect to longer ones. This finding is consistent with recent experiments which showed a greater thermal stability of shorter polyynes absorbed on Ag nanoparticles [31]. The interpretation of the experimental observations are supported by density functional theory calculations of the Raman response of silver end capped polyynes (Ag-$C_N$-Ag, $6 \leq N \leq 20$) which have been used as simple models able to mimic the remarkable chemical SERS effect observed for these systems [28, 30].

2. Experimental

*2.1 Polyyne production*

Polyynes were produced by electric arc discharge between two graphite electrodes submerged in 100 ml of methanol in a three-necked round bottomed flask. The electric arc was conducted under the usual conditions of 10 A and electrodes arranged in a "V" geometry with external cooling in a water/ice bath [14, 32, 33]. The arc was prolonged for 30 min. and then the crude mixture was subjected to high performance liquid chromatography (HPLC) analysis, after filtration through a polyvinylidenefluoride (PVDF) filter [34].

Individual polyynes separated by the HPLC column were identified both on the basis of their retention time and their electronic absorption spectra [34]. Absorption spectra of C8, C12 and C16 polyynes reported in Figure 1 show characteristic π-π* transitions in the UV spectral region (200 – 350 nm range).

The concentration of each species was measured on the basis of the absorbance of their most intense peak in the electronic absorption spectra by using the Lambert-Beer law and the molar extinction coefficients reported in the literature [35]. Typical distributions of chain lengths H-$C_N$-H range from N = 6 up to N = 16 with a relative abundance maximum for N = 8 [28].

*2.2 Preparation of silver colloids*

$AgNO_3$ (99%) and trisodium citrate (98%) from Aldrich chemicals were used without further purification. A modified Lee and Meisel [36] procedure was implemented to obtain highly concentrated silver colloids. 200 mg of $AgNO_3$ were dissolved in 500 ml of distilled water and brought to boiling. 20 ml of a 3% trisodium citrate solution were added and maintained at boiling until the color turned to orange. The solution was then cooled to room temperature and placed in a sealed ampoule for about 1 month. After this time, highly concentrated colloids can be extracted from the bottom of the flask.

Plasmon resonance of silver colloids has been investigated by UV-Vis absorption spectroscopy using a V-570 Jasco spectrophotometer. Scanning transmission electron microscopy (STEM) images of silver nanoparticles were taken with a Zeiss Supra 40 field emission SEM equipped with a STEM detection module. The STEM sample was prepared by drying a droplet of colloidal solution on a TEM grid covered by a thin carbon layer.

*2.3 SERS experiments*

Silver colloids were added to the polyyne solution in $CH_3OH$ in order to perform SERS measurements. SERS spectra were recorded with a Nicolet NXR9650 FT-Raman (resolution 4 cm$^{-1}$) equipped with a InGaAs detector and a Nd:YVO$_4$ laser providing a 1064 nm excitation line. SERS spectra have been recorded on aqueous colloidal suspension containing silver nanoparticles, methanol (used as reference) and polyynes (sample molecules). The polyyne solution (~ $10^{-5}$ M) has been mixed with the silver colloid keeping a 1 to 1 volume ratio. For direct comparison all the spectra have been recorded with the same experimental conditions (backscattering geometry, laser power about 0.3 W at the sample, collection time 30 sec.). We estimated a SERS intensification between $10^5$ and $10^6$ [28].

**3. Theory**
Density functional theory calculations of the off-resonance Raman response have been carried out using the pure Perdew-Becke-Ernzerhof (PBE) exchange and correlation functional [37]. We have selected the 6-311G** basis set for carbon and hydrogen atoms and the 3-21G* basis set for silver atoms. The theoretical method considered in this work employs the same basis sets used in a previous study [28] but the exchange-correlation functional PBE (widely used among generalized gradient corrected functionals) instead of BPW91 [38]. The comparison between these two theoretical approaches shows that the differences are minor and do not imply changes in the interpretation of the data. DFT calculations provide helpful information which nevertheless has to be considered with some care. In fact, it has been recently pointed out that while DFT can correctly predict the observed trend of the Raman response of polyynes, it is not able to account *quantitatively* for the observed red shifts of the strong Raman lines of hydrogen capped polyynes with increasing chain length. Suitable scaling procedures have been introduced to overcome this limitation [39, 40]. This inaccuracy of DFT calculations is likely to be an issue also for silver end-capped polyynes, but to date the proposed scaling procedures have not been adapted to these polyynic systems.

**4. Results and discussion**
Polyynes interact with silver nanoparticles used as the SERS active medium as already reported in [28, 29, 31]. This can be directly observed with a color change when polyynes are mixed with colloidal solution and is evidenced by the strong modification of the UV-Vis-NIR absorption spectra (Fig. 2). As prepared SERS active silver colloids, consisting of both polyhedric-shaped (20-100 nm) and rod-shaped (of different length and 30-50 nm wide) Ag nanoparticles, show a plasmon resonance centered near 450 nm (Fig. 2-a). The distribution of size and shape of silver nanoparticles can account for the broadness (110-120 nm FWHM) of the observed plasmon peak. A red-shift and a further broadening of the plasmon resonance peak occur after mixing with polyynes. Plasmon resonance can be strongly affected by aggregation since a dipole coupling takes place when isolated nanoparticles are brought at a close distance [41]. Aggregation is usually induced by modifying the ionic strength of the solution (hence the nanoparticle electric double layer), for instance by adding NaCl or by introducing molecules interacting wih the nanoparticle surface. The comparison between these two aggregation methods (Fig. 2-b) clearly reveals that polyynes induce Ag colloids aggregation. In many cases a broadening of the absorption spectrum is a desired effect since it allows to perform SERS measurements using excitation lines far from the intrinsic plasmon resonance (even in the near-IR) of as prepared silver colloids.

Interaction of polyynes with Ag nanoparticles has also a role in determining the time evolution of the SERS spectra. Fig. 3 reports SERS spectra at different times after mixing (between 2 and 65 minutes) where the methanol peak at 1020 cm$^{-1}$ has been used for normalization. Two features are observed: one at 1800 – 2200 cm$^{-1}$, related to CC stretching vibrations of sp carbon [28], while the other at 1000 – 1700 cm$^{-1}$ can be largely ascribed to sp$^2$ carbon containing molecular species (here named sp$^2$ for simplicity). Visual inspection of the sample during time after the mixing of the colloid and the polyyne methanol solution provides useful information: (i) at early times we observe a change in the colour of the colloid which turns dark due to aggregation and therefore plasmon resonance broadening (see Fig. 2-a); (ii) as the aggregates keep growing, the colloidal suspension becomes unstable and a precipitate is observed on the bottom of the sample tube. This latter effect has been already observed [42] and is responsible for a decrease of the SERS signal which eventually can disappear when precipitation is complete.

An overall SERS signal increase at early times after mixing (2-7 minutes) is followed by a later decrease (>10 minutes) as shown in Fig.4-a. This behavior is due to silver colloid aggregation and polyynes adsorption and to precipitation of large aggregates from the colloidal solution. These phenomena have opposite effects on the overall SERS signal. In any case, the two effects have different characteristic times. Indeed initial aggregation and adsorption of polyynes on silver nanoparticles enhances the SERS signal by providing better resonance conditions with respect to the 1064 nm excitation line and a higher number of hot spots and molecules on the silver surface [43]. Then precipitation of the aggregates weakens the SERS signal due to the decrease of the number of SERS active aggregates in the probed volume. Moreover evolution of polyynes (i.e. sp-sp$^2$ conversion and variation in the chain length distribution) must be also taken into account.

In addition to the physical phenomena associated to the SERS effect the spectra show that chemical effects also occur upon interaction of polyynes with Ag nanoparticles. By comparing Raman and SERS spectra we observed shifts of sp features and appearance of new peaks revealing a chemical effect [28]. We pointed our attention on the sp → sp$^2$ conversion. To this aim, the intensity of the sp and sp$^2$ features have been normalized with respect to the total intensity. The time evolution of sp and sp$^2$ relative intensity, reported in Fig3-b, reveals a decrease of the sp feature with a parallel increase of the sp$^2$ one. Both time evolutions seem to follow a simple exponential law:

$I(t) = A \cdot \exp(-t/\tau) + I_0$

with the same decay time of about 12 minutes. It has to be noticed that this fit is used here to analyze the relative evolution of sp and $sp^2$ signal intensities, while a quantitative evaluation of sp and $sp^2$ content is extremely difficult without considering the actual SERS cross sections. The observed behavior could be due to degradation of sp phase into $sp^2$ phase due to chain cross linking [44], in agreement with the tendency of sp phase to undergo transition towards the more stable $sp^2$ phase [13].

Once the overall sp-$sp^2$ evolution has been outlined, we focused on the sp band which looks structured in distinct peaks, as already reported [28]. Since SERS spectra have been taken with 1064 nm excitation and typical π-π* excitations of polyynes lie in the UV (see Fig. 1) [14], in a first approximation one can disregard resonance effects in analyzing these data. Fig. 5 presents the same SERS spectra of Fig. 3 in a reduced spectral range (1750-2300 $cm^{-1}$). In order to exclude the already discussed changes of the total sp intensity, all spectra have been normalized with respect to the total sp signal. This allows to follow the relative contributions of the main three peaks constituting the sp band, namely $p_1$ at 1910 $cm^{-1}$, $p_2$ at 2020 $cm^{-1}$, $p_3$ at 2115 $cm^{-1}$. The evolution as a function of time of the normalized intensities of the peaks in Fig. 5 (namely $p_1$/sp, $p_2$/sp, $p_3$/sp) is shown in Fig.6 following the same simple exponential law reported above. $p_1$ and $p_2$ show a relative decrease while $p_3$ increases (values are reported in Table 1).

DFT calculations allowed to interpret the three contributions $p_1$, $p_2$, $p_3$, as due to signals produced by chains of various length. The chemical interaction between the colloidal silver nanoparticles and the hydrogen capped polyynes [29] has been modeled considering linear carbon chains capped at both ends by silver atoms [28]. Simulated Raman spectra are dominated by just one (in some cases two) bands attributed to collective CC stretching vibrations (see Fig. 5). The inspection of the associated nuclear displacements has shown that these vibrations, similarly to the case of hydrogen capped polyynes, can be associated to longitudinal optical phonons of the corresponding infinite polyyne (for a detailed discussion of the vibrational dynamics and Raman response of carbon linear chains see [6, 26, 39, 45-48]). The analysis of the CC bond lengths and Mulliken charges of hydrogen and silver capped polyynes (H-$C_8$-H, H-$C_8$-Ag and Ag-$C_8$-Ag) is reported in Figure 7. The Bond Length Alternation (BLA) is smaller for the silver capped Ag-$C_8$-Ag polyyne with respect to the hydrogen capped one (H-$C_8$-H). In particular, the external triple bonds (namely positions -3 and 3) are longer in Ag-$C_8$-Ag than in H-$C_8$-H while H-$C_8$-Ag shows an intermediate behavior. This change in BLA is consistent with a charge transfer process involving the silver atoms and the neighbouring carbon atoms as confirmed by the analysis of Mulliken charges (see Figure 7-b). Figure 7-b show that the presence of silver significantly affects the atomic charges along the carbon chain with respect to the hydrogen capped poyyne H-$C_8$-H. Silver atoms exhibit a positive charge and the closest carbon atoms (namely $C_3$ and $C_8$) have a negative charge. This induces a significant perturbation on the electronic structure of the carbon chain and is accompanied by a change in the vibrational frequencies while passing from hydrogen to silver capped polyynes [28]. The terminal CC triple bonds (positions -3 and 3 of Figure 7-a) are the more affected by the presence of silver atoms and consistently the charges of the atoms defining these CC bonds ($C_2$-$C_3$; $C_8$-$C_9$) are markedly different with respect to the hydrogen capped case.

These results indicate that, because of the dispersion of the strong Raman band with length, $p_1$, $p_2$, $p_3$, are related to the convolution of the signals produced by contributions of longer, intermediate and shorter carbon chains, respectively (see Fig. 5). This attribution is indeed confirmed by SERS experiments carried out on hydrogen capped polyynes of selected chain lengths [30], even though a precise assignment to specific chain lengths in not possible, due to the difference between the SERS active substrates (silver colloids with respect to silver islands films) and the essential role of the substrate-analyte interaction in SERS. Anyway, we can ascribe the trend of $p_1$, $p_2$ and $p_3$ as due to a relative decrease of longer polyynes with respect to the shorter ones. To determine single chain length fraction of polyynes in the sample one should consider also the SERS intensity of each chain which is expected to be sensibly chain length dependent as suggested by our simulations (Fig. 5).

The experimental and theoretical data here presented suggest an overall trend towards conversion of sp phase into the more stable $sp^2$. Within this evolution of the sp band, shorter chains seem to be more stable than longer ones. In fact the contribution of shorter chains to the sp band, initially of about 10%, increases up to more than 30% after 65 minutes as a result of the decrease of the contribution of medium and longer chains (from 45% to 40% and 30% for longer and medium chains, respectively). According to the theoretical data reported in Fig.4, it is likely that after 65 minutes shorter chains represent the major fraction in the sample since the cross section is substantially lower than that of longer chains. The higher stability of shorter chains (6-10 atoms in the as prepared sample) is somehow in agreement with the observation of $C_8H_2$ as the most abundant species in solution while longer chains are more difficult to be produced [23]. Also a low stability of sp chains with conversion to $sp^2$ was observed in other sp carbon systems [44]. For instance high energy release is observed when isolated sp chains embedded in solid inert gas matrices interact to form $sp^2$ network [13, 49]; formation of graphitic nano-domains induced by thermal treatments in sp-$sp^2$ amorphous carbon films deposited by low energy cluster beam deposition has been also reported [10].

**5. Conclusions**

The SERS technique has been successfully used to investigate the interaction of sp carbon chains (polyynes) with silver colloids (the SERS active medium) as a function of time. Thanks to the high enhancement achieved in SERS, we have been able to follow the time evolution in the 2-65 minutes range of the overall sp-$sp^2$ conversion and of the contribution to the sp band of chains with different lengths. When a solution of polyynes in methanol is mixed with silver nanoparticles several reactions take place due to the aggregation induced by the strong interaction of polyynes with silver. After proper spectra normalization in order to exclude other effects affecting the SERS signal we can obtain information on both the $sp^2$/sp ratio and the internal ratios within the sp band of the SERS signal. We have shown that under our operative conditions (methanol solution of polyynes mixed with silver aqueous colloids) the $sp^2$/sp ratio increases and the chain length distribution of polyynes converges towards more stable shorter chains. These results, with the support of DFT calculation of the Raman modes of silver end capped polyynes, suggest the hypothesis of cross-linking processes is more effective in medium and longer chains. Even if additional studies and experiments are still necessary to further clarify the interaction of polyynes with silver nanoparticles, we demonstrated that SERS is a powerful technique to investigate the structure and the stability of sp carbon chains also giving access to the time evolution of complex processes involving such form of carbon structures.

**Acknowledgments**

This work has been partly supported by grants from the Italian Ministry of Education, University and Research through FIRB projects *"Molecular compounds and hybrid nanostructured materials with resonant and non resonant optical properties for photonic devices"* (RBNE033KMA) and *"Carbon based micro and nano structures"* (RBNE019NKS), by project PRIN *"Molecular materials and nanostructures for photonics and nanophotonics"* (2004033197) and by FlagProject *"ProLife mobilità sostenibile"* funded by the Milano city administration. The authors acknowledge A. Bonetti and S. Salvatore for the contribution given during their undergraduate thesis project.

# Captions to figures and tables

**Fig. 1.** Absorption spectra of as prepared polyynes (C8, C12 and C16) in methanol solution.

**Fig. 2.** (a): Extinction (absorption + scattering) spectra of silver colloids measured before and after aggregation induced by mixing colloids with polyynes in methanol. (b): Extinction (absorption + scattering) spectra of silver colloids measured at increasing NaCl concentrations showing the aggregation of nanoparticles (see text). STEM image of as prepared Ag nanoparticles used as the SERS active medium is also reported.

**Fig. 3.** Time evolution of SERS spectra recorded on a colloidal solution of silver nanoparticles and polyynes in methanol. Time is measured after the mixing of polyynes with the concentrated silver colloid. The arrow indicates a methanol peak at 1020 cm$^{-1}$.

**Fig. 4.** (a): Evolution with time of the integrated SERS signal of sp and sp$^2$ from the spectra reported in Fig. 3. The grey line is a guide to the eye. (b): evolution of the sp and sp$^2$ signal normalized with respect to the total SERS signal. The grey lines show data fit by a simple exponential decay law.

**Fig. 5.** Time evolution of SERS spectra normalized to the integral over the frequency range 1750 - 2300 cm$^{-1}$. Bars represent calculated frequency and intensity from first-principles calculations of off-resonance Raman response of silver end capped polyynes Ag-C$_N$-Ag ($6 \leq N \leq 20$) (see text).

**Fig. 6.** Time evolution of the normalized SERS intensity of the three main features p$_1$/sp, p$_2$/sp, p$_3$/sp according to the three gaussians fit of Fig. 5.

**Fig. 7.** Bond length alternation (a) and Mulliken charges (b) for C8 hydrogen and silver capped polyynes (H-C8-H, H-C8-Ag, Ag-C8-Ag). Data are obtained from DFT calculations carried out with PBEPBE functional, 6-311G** basis set (C and H) and 3-21G* basis set (Ag).

**Table 1**: Fitting values for simple exponential decay fit of SERS intensity of p$_1$, p$_2$, p$_3$ peaks in the sp band reported in Fig. 5

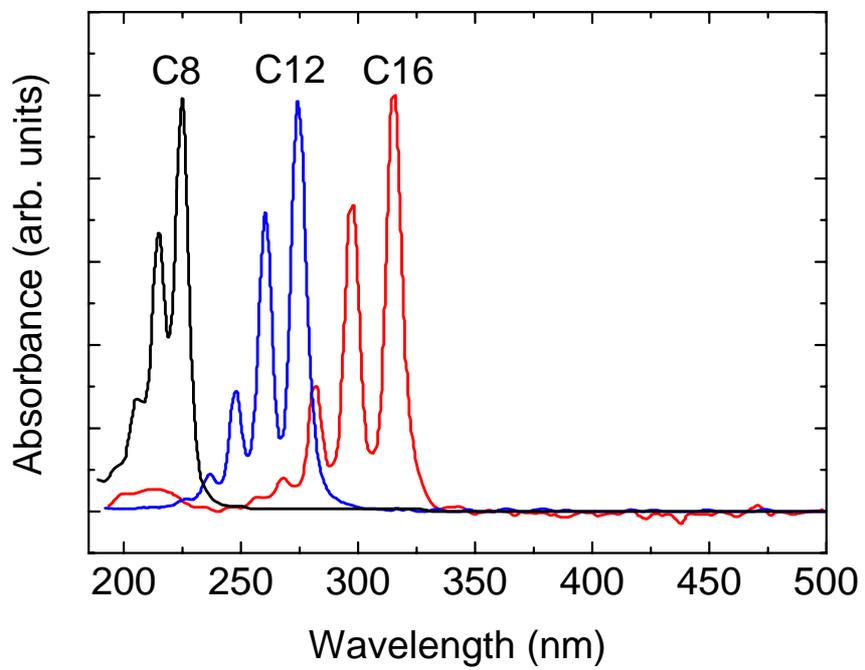

Fig.1

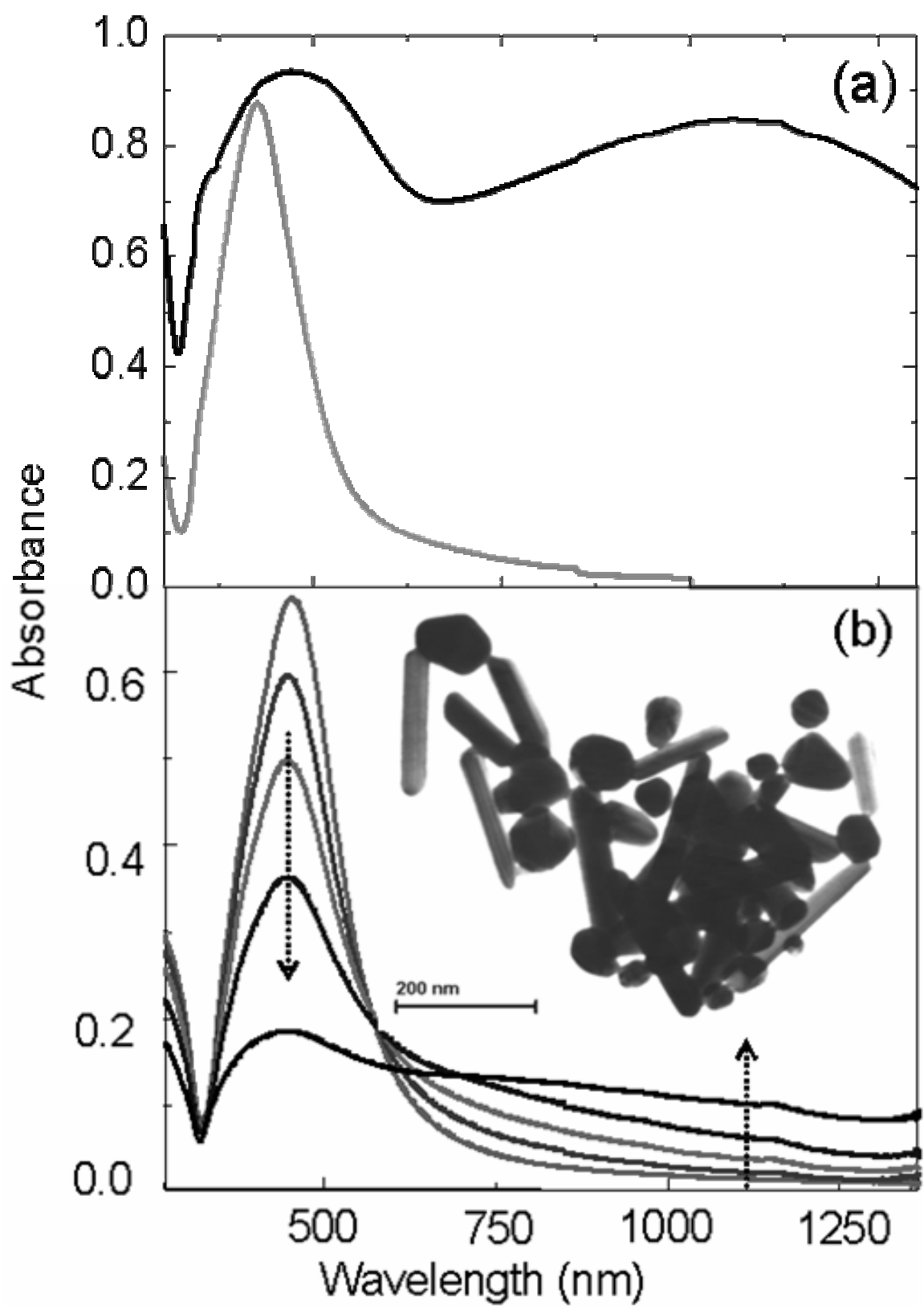

Fig.2

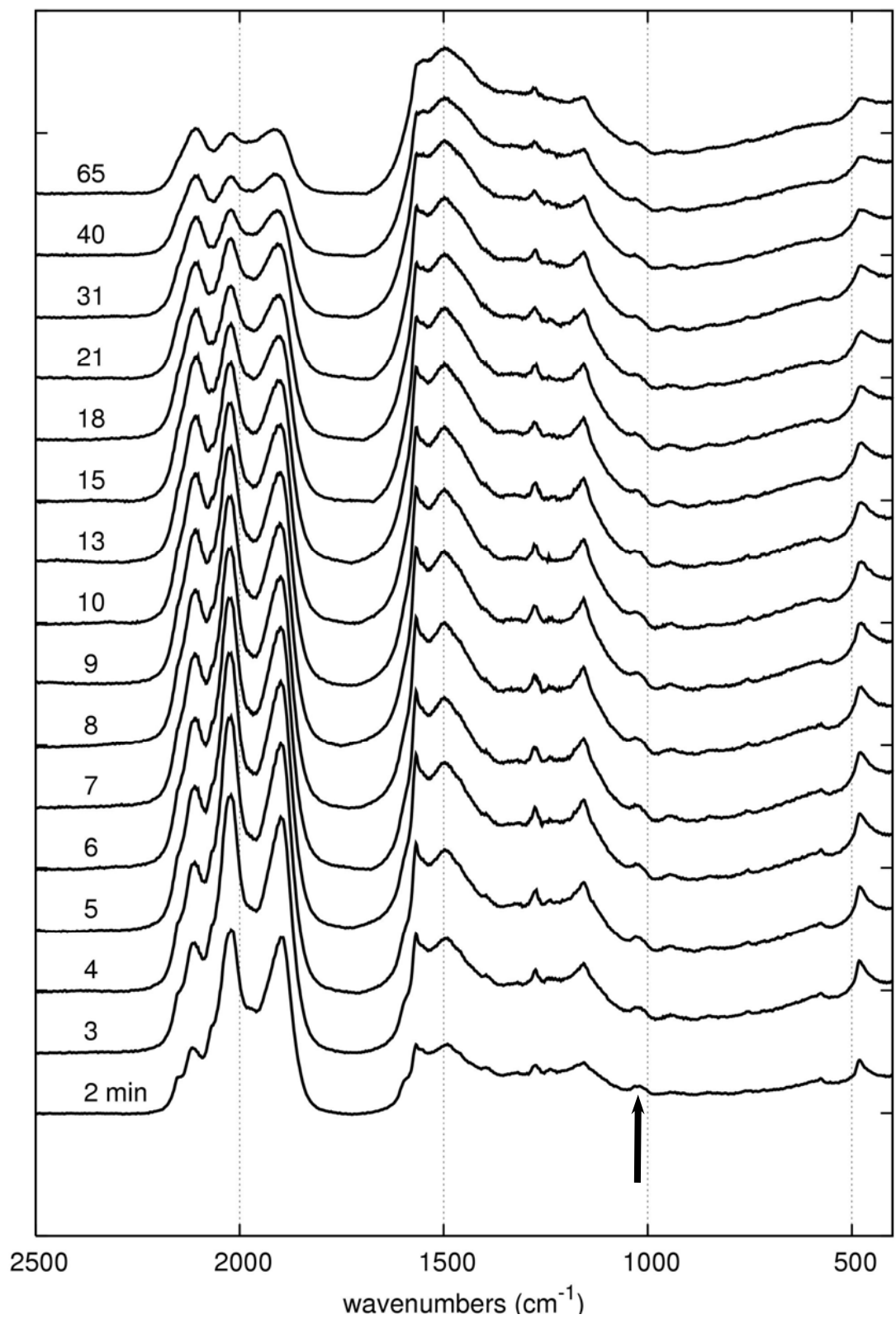

Fig.3

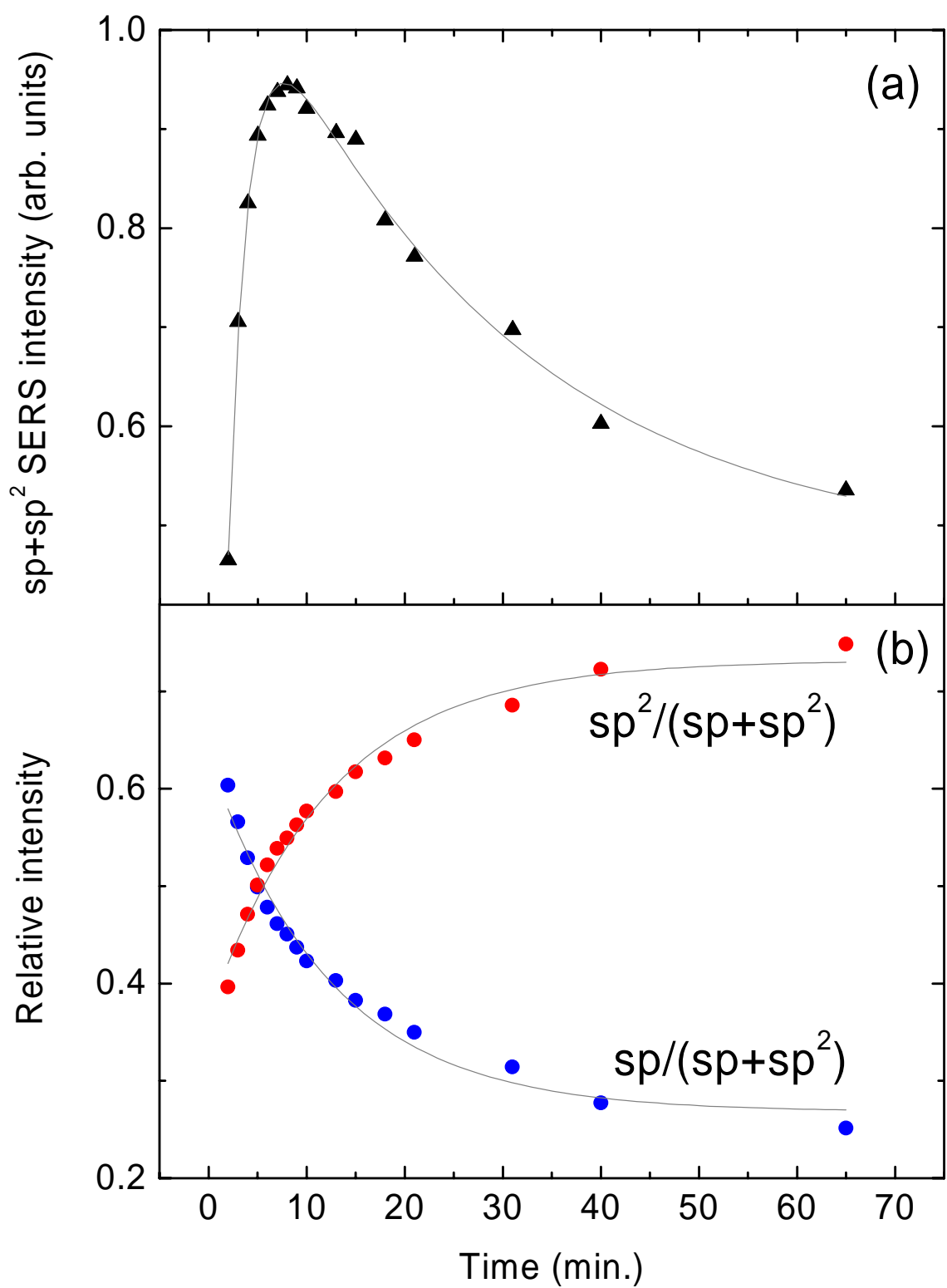

Fig.4

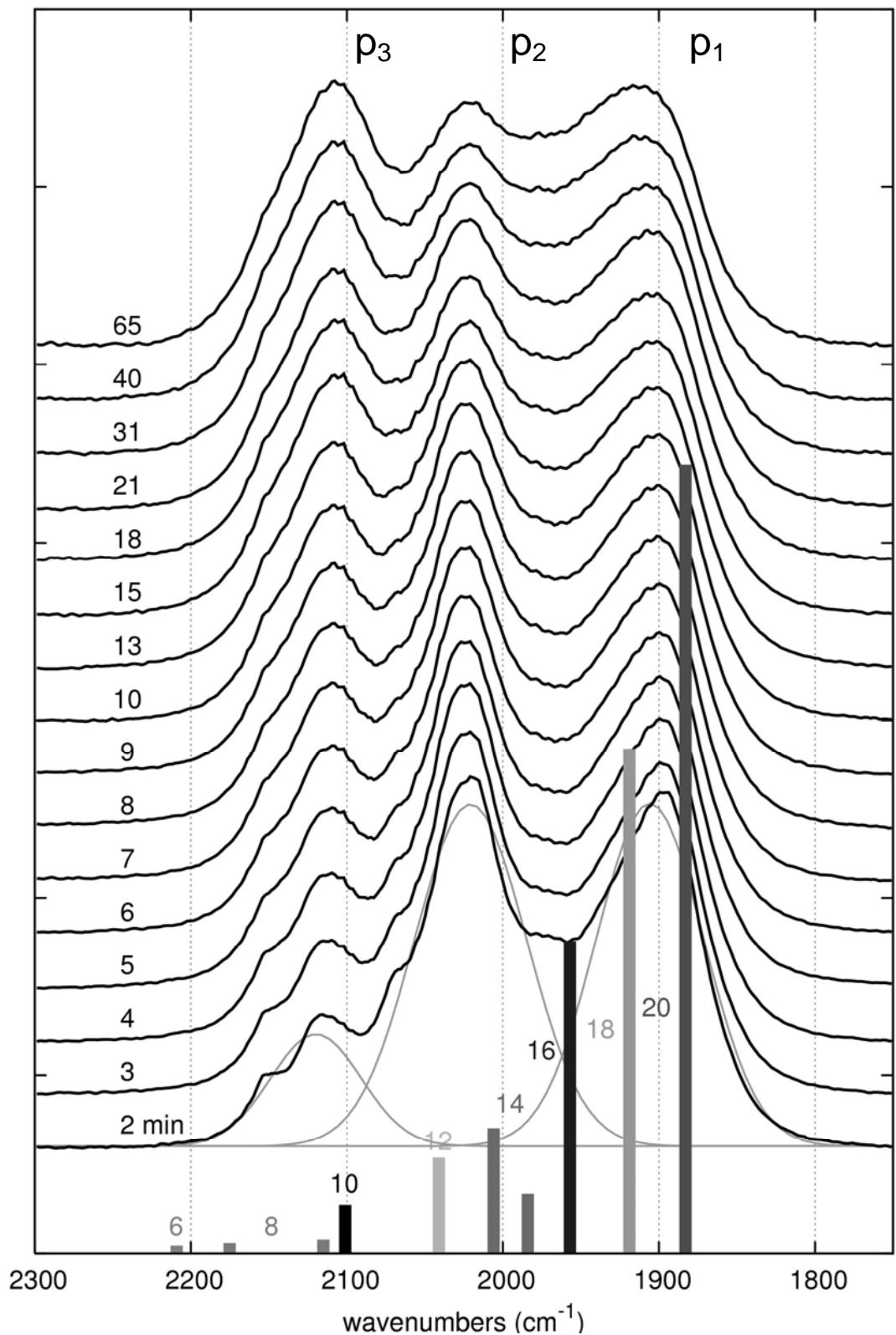

Fig.5

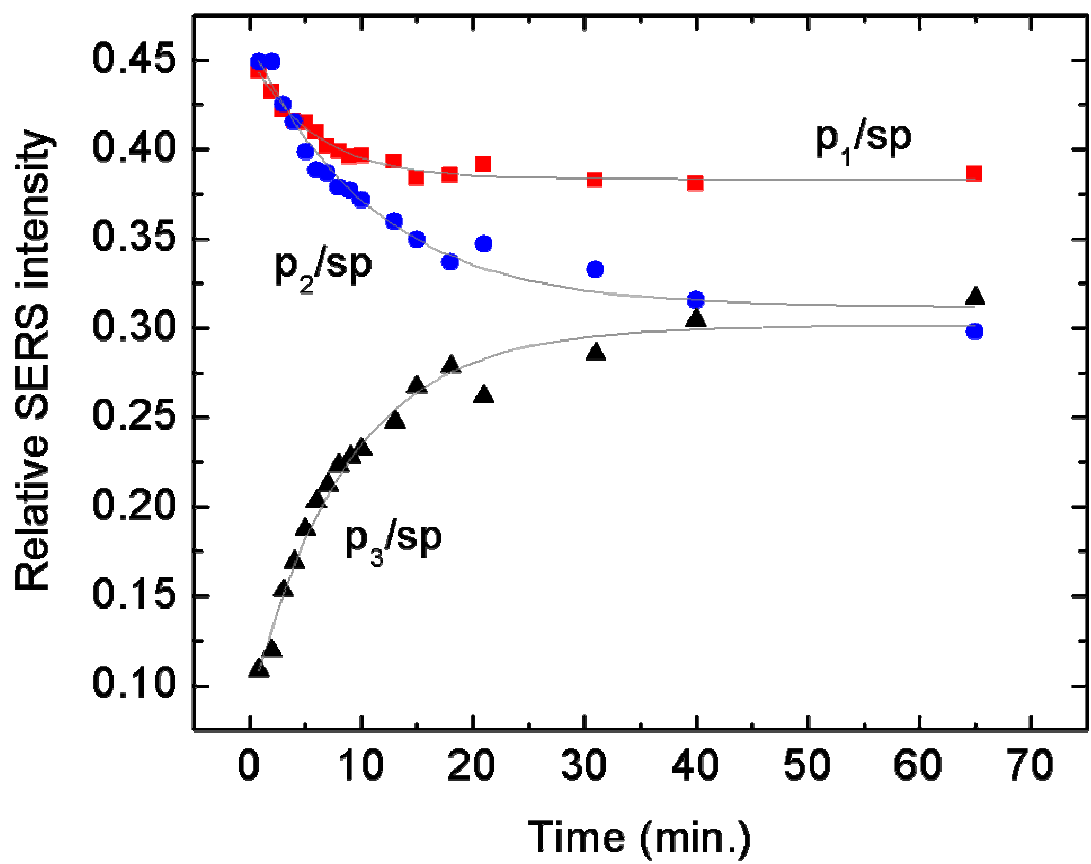

Fig.6

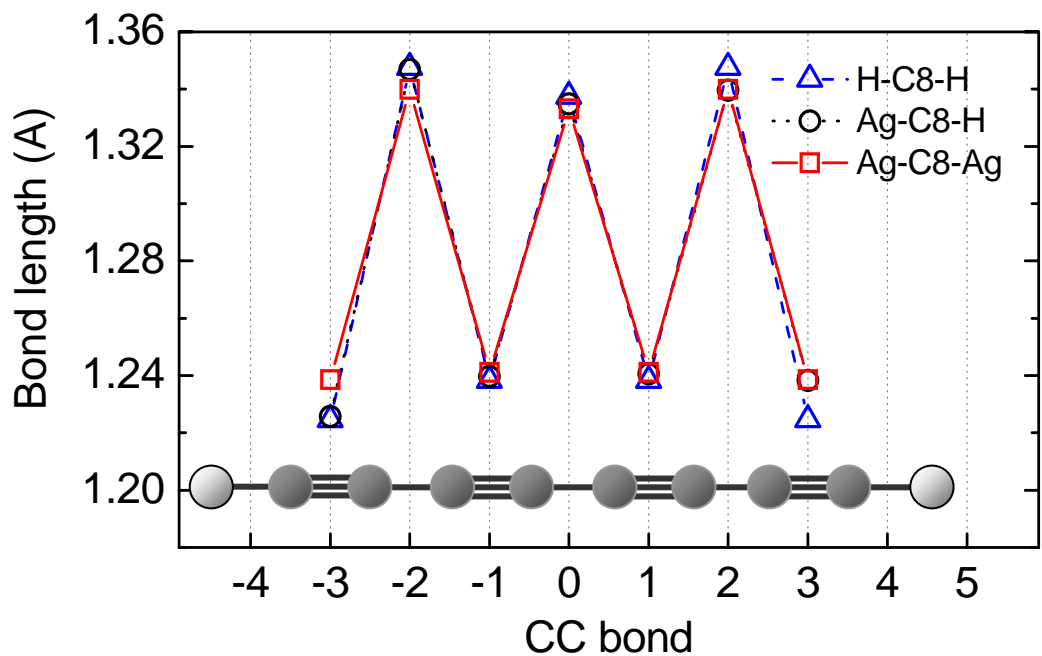

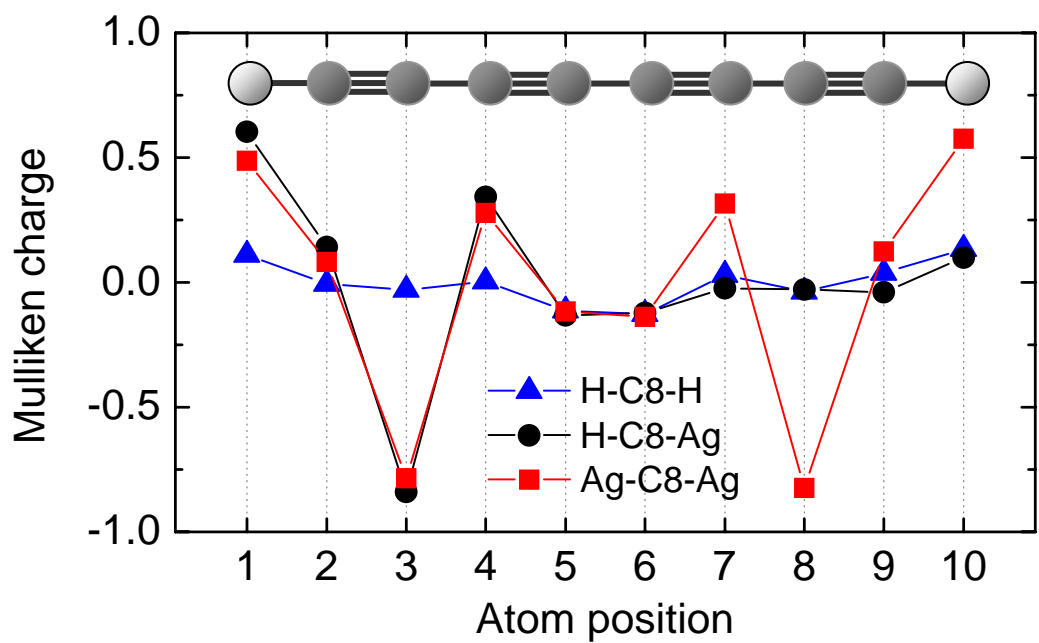

Fig.7